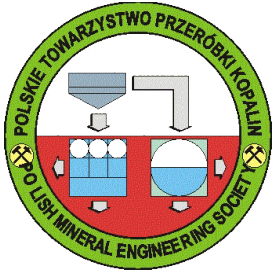

# Extending the simulations of intermediate-mass black hole mass measurements to Virgo Cluster using ELT/HARMONI high resolution integral-field stellar kinematics

*Hai N NGO[1*], Dieu D. NGUYEN[2], Truong N. NGUYEN[1], Trung H DANG[1], Tien H.T. HO[1]*

[1]Faculty of Physisc – Engineering Physics, University of Science, Vietnam National University - Ho Chi Minh City, Ho Chi Minh City, Vietnam.

HNN: Email: hai10hoalk@gmail.com

DDN: nddieuphys@gmail.com

TNN: nntruong@hcmus.edu.vn

THD: dhtrung@hcmus.edu.vn

THTH: htien2808@gmail.com

ORCID iDs:

Hai N Ngo: https://orcid.org/0009-0006-5852-4538

Dieu D. Nguyen: https://orcid.org/0000-0002-5678-1008

Truong N. Nguyen: https://orcid.org/0009-0005-1191-7775

Trung H. Dang: https://orcid.org/0000-0002-8089-9380

Tien H.T. Ho: https://orcid.org/0009-0005-8845-9725

*Abstract*

*The firm co-existence of intermediate-mass black holes (IMBHs, $M_{BH} \approx 10^3 - 10^6 M_\odot$) in nuclear star clusters (NSCs) remains uncertain because the limited number of verified instances within the local Universe, limited within 3.5 Mpc. They are crucial for our understanding about the formation and evolution of supermassive black holes (SMBHs). The upcoming Extremely Large Telescope (ELT) promises to revolutionize the detection of these mysterious objects. In this study, we simulated the kinematics of an IMBH within the nuclear star cluster of VCC 1861, one of the faintest galaxies in the Virgo Cluster. Using Jeans Anisotropic Modeling (JAM) and stellar density profiles derived from Hubble Space Telescope (HST) data and the HSIM program, we created mock High Angular Resolution Monolithic Optical and Near-infrared Integral field spectrograph (HARMONI) observations for the ELT. We then extract stellar kinematics from* *these mock data and recover the BH mass using the JAM model with Markov Chain Monte Carlo simulation method. Our results demonstrate the ELT's capability to detect IMBHs with masses comprising 5% of the NSC's mass at the distance of the Virgo Cluster.*

**1. Introduction**

The intermediate-mass black holes (IMBHs, with a mass range of $10^2 \lesssim M_{BH} \lesssim 10^6 M_\odot$) existence in below-Milky Way mass galaxies ($M_\star \lesssim 5 \times 10^{10} M_\odot$) is important to our knowledge of the origins of SMBHs (SMBHs, $M_{BH} \gtrsim 10^6 M_\odot$) in early Universe. Currently, we knew that the stellar mass black holes (BHs, $M_{BH} \lesssim 10^2 M_\odot$) is massive star remnants [1]. In contrast, the origins of SMBHs at the centers of larger galaxies remain unclear despite their ubiquitous presence. Observations of luminous quasar from the reionization epoch (at $z \gtrsim 6$) reveal some large SMBHs with mass $M_{BH} \gtrsim 10^9 M_\odot$[2]–[6]. These discoveries suggests that SMBHs could not have evolved from normal stellar-mass BHs in this short time. This implies some hypothesized mechanisms for the origins of SMBHs involve starting as IMBH "seeds" which then eventually grow to their current SMBHs we observed today in massive galaxies.

There are three theoretical channels for IMBH formation: (i) from remnants of massive stars (pop-III) in the early Universe [8], [9], (ii) via gravitational runaway in dense stellar system [10]–[13], and (iii) direct collapse from gas [14], [15]. All these hypotheses require IMBHs as initial seeds formed around a quarter billion years from the Big Bang and then grew up inside the galaxies. On the other hand, if these IMBHs did not evolve into SMBHs, they should be found in low-mass stellar systems (e.g. dwarf galaxies [16]–[18] or globular clusters (GCs) [19]–[22]) due to a lack of material for accretion. By examining the relationships between central BH mass and host galaxy (e.g. velocity dispersion ($\sigma_\star$) or galaxy stellar mass ($M_\star$)) as well as the occupation fraction of IMBH inside low-mass galaxies, we can identify the dominant formation process.

However, strong observational evidence for such IMBH are absent from BH mass spectrum. Some extragalactic surveys detect no IMBHs within Local Group[23], [24]. Despite numerous multi-wavelengths surveys (e.g X-ray [25]–[27], Optical[28]–[30], near-infrared, radio[31]) with many techniques (e.g. dynamics, broad Balmer lines, coronal lines) aiming to find these enigmatic objects, leading their existence remains uncertain. However, some recent observations reveal strongest IMBH candidates in Local Volume dwarf galaxies (e.g. NGC 3621[32], NGC 4395[33], NGC 5206, NGC 5102, NGC 205 [34], [35], NGC 404 [36]–[38] ) and nearby GCs [22], [39], [40]. The main challenge in detecting IMBHs is the lack of facility resolution. The existence of IMBHs can be detected via their effects on their vicinity within their radius of sphere of influence $R_{soi}$. This radius is given by $R_{soi} = GM_{BH}/\sigma_\star^2$, where $G$ is the gravitational constant, $M_{BH}$ is black hole mass, and $\sigma_\star$ is the stellar velocity dispersion. For a typical IMBH in Local Group ($D \approx 3.5\ Mpc$) with mass of $M_{BH} = 10^5 M_\odot$ and $\sigma_\star = 30\ kms^{-1}$, then $R_{soi} \approx 0.028$ arcsec. This is below the resolution of all current telescopes, even with adaptive optics support such as the Very Large Telescope (VLT, FWHM$_{PSF}$ = 0.05 arcsec) or Gemini (FWHM$_{PSF}$ = 0.1 arcsec). We certainly can question whether the current absence of strong observational evidence for IMBHs suggests their natural non-existence or a result of our instrumental limitations.

In the coming years, Extremely Large Telescope (ELT) with 39-meter mirror will offer sufficient spatial resolution and is expected to be capable of resolving the Rsoi of IMBHs. In addition, it is equipped with High Angular Resolution Monolithic Optical and Near-infrared Integral (HARMONI) field spectrograph. It can achieve the spectral resolution of σ$_{instr}$ = 6 - 18 km.s$^{-1}$, allow us to measure IMBH mass in the nearby galaxies within 10 Mpc better than current facilities [41].

In this work, we aim to investigate the ability of ELT/HARMONI in IMBH mass measurement with Virgo cluster with distance of 16.5 Mpc. We introduced VCC 1861 as a typical dwarf galaxy, which serves as the target for our simulation (Section 2). To do that, we generated four mock IFS data with two different BH mass scenarios and extracted their kinematics in Section 3. Then we extracted kinematic information and recovery BH mass values using Bayesian inference in Section 4. Finally, we summarized these findings in Section 5.

## 2. Simulated Virgo cluster target: VCC 1861

Virgo Cluster is the one of the largest galaxy clusters at North hemisphere with ≈ 2,000 members [42]. It hosts numerous dwarf galaxies which predicted to harbor IMBHs [43], [44]. With its diversity in galaxy demographics [45], it provides perfect targets for studying galaxy evolution.

VCC 1861 (or IC 3652) is a typical dwarf elliptical galaxy (dE) in Virgo Cluster, locates at $12^h40^m58.58^s - 11^d11^m04.23^s$. It has estimated total stellar mass of $M_{\star,gal} \approx 1.26 - 3.55 \times 10^9 M_\odot$ [46]–[48] with an effective radius in *r*-band of $R_{e,gal} = 20.1$ arcsec [49]. This face-on galaxy displays quite round and regular isophotes, with a mean ellipticity $\langle \varepsilon \rangle = 0.03$ [50]. VCC 1861 contains a bright NSC with total magnitude in *g*-band of 20 mag [51] and mass of $M_{\star,NSC} = 5.2 \times 10^6 M_\odot$ estimated from spectral energy distribution fitting with Keck observations [46]. Spectroscopic studies indicate VCC 1861 has an average age of $Age_{gal} \approx 6.31$ Gyr and a younger NSC with $Age_{NSC} \approx 5.24$ Gyr [46]. The galaxy has a line-of-sight (LOS) rotational velocity of $v \approx 5 km.s^{-1}$ and $\sigma_\star \approx 28 km.s^{-1}$ from William Herschel Telescope and Keck data [47], [48].

## 3. Methodology

In this section, we simulated mock observations using the HARMONI instrument of the ELT. We generated the mock IFS data cubes with the HARMONI Simulation (HSIM[a]) program [52]. To achieve this, we first built a galaxy mass model based on Hubble Space Telescope (*HST*) images (Section 3.1). We then constructed a noiseless IFS cubes using this mass model, combined with a kinematic profile derived from Jeans Anisotropic Modeling (JAM) and a Stellar Population Synthesis (SPS) spectrum (Section 3.2).

### 3.1. Galaxy mass model

We constructed the galaxy stellar mass model using data from from HST/ACS Wide Field Channel (WFC) image downloaded from Hubble Legacy Archive. It is a part of Virgo cluster survey by [51]. This image is observed from three frames with total exposure time of 1210s. For our analysis, we generated the *HST*/WFC F850LP point spread function (PSF) using the Tiny Tim package [53], [54]. To replicate the real observational conditions, we created three separate PSFs corresponding to the three exposure frames. These PSFs were convolved into a final a pixel size of 0.05″ using the Drizzlepac/AstroDrizzle software [55]. The resulting PSF model has a full width at half maximum (FWHM) of 0″.11

---

[a] https://github.com/HARMONI-ELT/HSIM

We extracted the radial surface brightness profile using from Image Reduction and Analysis Facility (IRAF) program [56]. Firstly, we calculated the total flux within ellipse annuli using "ellipse" task from IRAF. During this step, we deconvolved the image with the *HST*/WFPC2 F850LP PSF described previously. We then applied a zeropoint of 24.872, obtained from the ACS Zeropoints Calculator in the *acstools* package[b], to convert the flux profile from counts.s$^{-1}$ to mag.arcsec$^{-2}$.

Notably, our chosen ELT/HARMONI pixel size of 0.01" (spatial resolution), which is five times smaller than the given HST observation. We need to extrapolate deeply toward the center. To do that, we fit the surface brightness profile with the core-Sersic [57], [58] and Sersic [59] function using non-linear least-squares method via MPFIT package [60]. As assumed by [41], the stellar mass distribution in NSC can be described using core-Sersic with the power law index of $\gamma = 0.1$. **Figure 1** shows the surface brightness profile and the best-fitting core-Sersic and Sersic funtion.

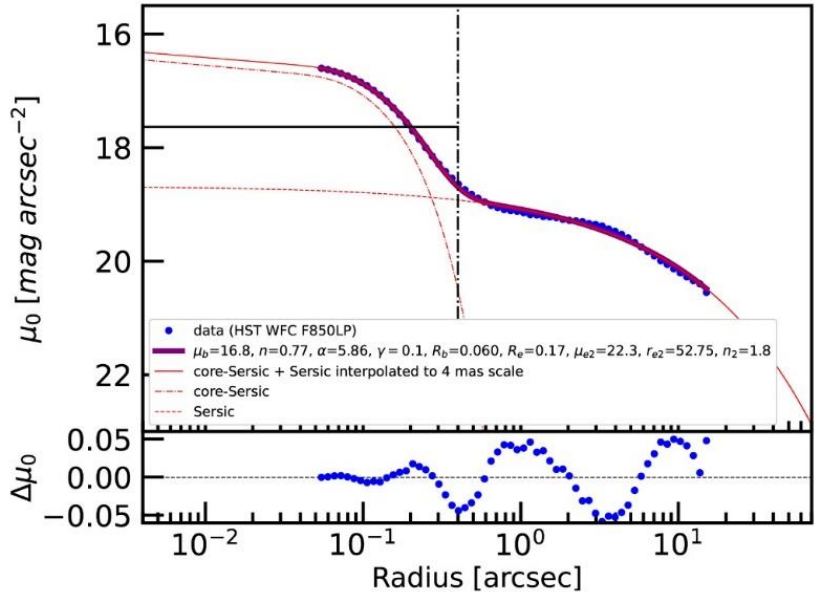

**Figure 1.** The surface brightness profile obtained from HST ACS/WFC image

We assumed the NSC component is perfect spherical with $\epsilon = 0$ and disk component has a mean ellipticity $\langle \epsilon \rangle = 0.03$. We used this information to convert 1D MGE to 2D MGE with $q = b/a = 1 - \epsilon$. The final MGE model is presented in **Table 1**, including: (column 1) central intensity in logarithmic scale ($log\ I'$), (column 2) the width of the Gaussian, and (column 3) the ratio between the minor and major axes.

---
[b] https://pypi.org/project/acstools/

Finally, we convert this luminosity density profile to mass density profile by apply a constant mass-to-light (M/L) ratio. We used an empirical relation between color index ($g - z$) and M/L [61] $M/L_z = a_z(g - z) + b_z$, where $a_z = 1.25$ and $b_z = -1.11$. We adopted the value of $g - z = 1.04$ obtained [51] and calculated $M/L_{F850LP} \approx M/L_z = 1.4\ (M_\odot/L_\odot)$. Our MGE mass model shows consistency with previous studies [46], [47], estimating a total stellar mass for the galaxy of $M_{\star,gal} = 2.8 \times 10^9 M_\odot$ and NSC mass of $M_{\star,NSC} = 5.2 \times 10^6 M_\odot$

We consider the BH as a point-mass at galaxy center. We simulated with two scenarios: without a black hole ($M_{BH}$ = 0) and contains a IMBH with mass of $M_{BH} \approx 5\% M_{NSC} = 2.5 \times 10^5 M_\odot$. This aims to compare and assess the capability of distinguishing the IMBH presence from the kinematic data.

Table 1. The Multi-Gaussian Expansion model obtained from HST/WFC 850LP image

| $logI'(L_\odot pc^{-2})$ (1) | $\sigma$ (arcsec) (2) | $q'$ (3) |
|---|---|---|
| 3.15 | 0.008 | 1.00 |
| 3.57 | 0.098 | 1.00 |
| 3.39 | 0.188 | 1.00 |
| 2.86 | 0.980 | 0.97 |
| 2.18 | 2.741 | 0.97 |
| 2.14 | 6.218 | 0.97 |
| 2.05 | 12.653 | 0.97 |
| 1.89 | 25.309 | 0.97 |
| 1.52 | 57.735 | 0.97 |

### 3.2. Mock HSIM IFS cube creating

For IFS cube creating, we used a high spectral resolution stellar population synthesis (SPS) spectra [62]. These spectra are constructed from the Model Atmospheres with a Radiative and Convective Scheme (MARCS) [63]. We selected it because its resolution ($R = \lambda/\Delta\lambda = 20,000$) is sufficient to match the ELT/HARMONI spectral capabilities. For our analysis, we adopted a Salpeter initial mass function (IMF), an NSC age of 5 Gyr [47], and solar metallicity. This template spectra is truncated from 1.4 – 2.5 μm, which can cover from H-high band to K-long band in our simulations.

We created the noiseless IFS cube with the same method in [41]. The galaxy's kinematic profile was constructed using Jeans Anisotropic Model (JAM). Because NSCs in dwarf galaxies tend to exhibit a significant rotation [34], [36], [64], we assumed an axisymmetric velocity ellipsoid with cylindrical alignment [65] $\sigma_z \neq \sigma_r = \sigma_\theta$.

We created two noiseless cubes: $M_{BH} = 0$ and $M_{BH} = 2.5 \times 10^5 M_\odot$ using mass model from Section 3.1. Other necessary parameters used for modeling are shown in Section 2. All JAM$_{cyl}$ model kinematic maps were constructed with assuming an average inclination $i = 44°$ and constant anisotropy parameter $\beta_z = (1 - v_z/v_R)^2 = 0.02$ [67]. They have the FoV of 0.8 × 0.8 arcsec and pixel size of 5 × 5 mas². This pixel size is smaller than the 10 × 10 arcsec² of ELT/HARMONI that we plan to use, allowing for later rebinning. We compute the rotational velocity $V$ and root-mean-square velocity $V_{rms}$ as first and second moment in JAM$_{cyl}$. Then the velocity dispersion is calculated by $\sigma_\star = \sqrt{V_{rms}^2 - V^2}$. We adopted the redshift $z = 0.0021$ from NASA/IPAC Extragalactic Database[c] (NED) by shifting the spectrum with a factor of $(1 + z)$.

With these assumptions, we generated the noiseless cubes by following steps:

(i) We constructed a Gaussian kinematic kernel with mean rotational velocity $V$ and velocity dispersion $\sigma_\star$ are calculated from JAM$_{cyl}$.

(ii) We assumed a stellar population (age of 5 Gyr, solar metallicity) and logarithmically rebinned the MARCS SPS spectra with velocity scale of 0.5 km.s$^{-1}$ to get a constant wavelength interval. We then acounted for redshift $z$ by a factor of $(1 + z)$.

(iii) We combined the spectrum from step (i) with the kinematic kernel in (ii), and then linearly interpolated to achive $\Delta\lambda \approx 0.02$ Angstrom.

(iv) We rebinned the spectrum to achieve a spectral resolution of at least half of the HARMONI spectral resolution ($\Delta\lambda \approx 0.5$ Angstrom). This ensures that no information is lost.

(v) The surface brightness is estimated from MGE model (Table 1). The elliptical shape of the galaxy was accounted for each pixel using the relation $r^2 = x^2 + (y/q)^2$, where $x$, $y$ is the positions of each spaxel, and $q$ is the ratio between minor and major axis in MGE model. This MGE model is then assigned for every pixel with unit of in erg.s$^{-1}$.cm$^{-2}$.Å$^{-1}$.arcsec$^{-2}$.

---

[c] https://ned.ipac.caltech.edu/

(vi) We saved the data into a 3-dimensional noiseless cube.

### 3.3. HSIM IFS output data cubes

HARMONI is a near-infrared integral-field spectrograph (IFS) instruments of ELT [68]. Assisted by an AO system, it can achieve the spatial resolution close to the diffraction limit which surpasses all previous capabilities. HARMONI can use some different AO modes, including single conjugate AO (SCAO), laser tomographic AO (LTAO), and no AO observations. It provides four different spatial resolution ($4 \times 4$ mas$^2$, $10 \times 10$ mas$^2$, $20 \times 20$ mas$^2$, and $30 \times 60$ mas$^2$) and three spectral resolutions ($R \approx 3{,}300; R_{medium} \approx 7{,}100; R_{high} \approx 17{,}300$).

We chose high-resolution gratings for precise kinematic measurement due to the small differences in the velocity profile. We simulated with two high-resolution gratings: H-high band ($\lambda 1.538 - 1.678 \, \mu m$) and K-long band ($\lambda 2.199 - 2.400 \, \mu m$). In these spectral ranges, there are some strong stellar atomic and molecular absorption lines (e.g. Mg, Fe, Si and CO absorption lines) which are used widely in kinematics extracting [71], [72].

We used noiseless cubes from Section 3.2 as the input for HSIM mock data simulations. The HSIM program truncated the spectral ranges of these noiseless cubes according to the respective gratings. We simulated multiple frames to mimic real data, with a total exposure time of 18 frames × 15 minutes = 4.5 hours. We used LTAO mode with a natural guiding star (magnitude in H-band of 17.5 mag) distant 30 arcsec from our target. The other parameters were set to default so mimic observational conditions for the Armazones site.

## 4. Result

### 4.1. Kinematic extracting

Firstly, we used the Voronoi binning method to merge adjacent pixels and achieve a signal-to-noise ratio $S/N \gtrsim 20$ for each bin using the vorbin package [73]. Next, we fitted the spectra of these bins to the MARCS spectral library using the Penalized PiXel-Fitting (pPXF) method [74]. The LOSVD was fitted using the Gauss-Hermite series. We accounted the continuum shape with default Legendre polynomials setting of degree = 4 and mdegree = 0. To improve the fitting process, the model is included 13 spectral templates that span a wide range of ages and metallicity in pPXF fitting process. We fitted the entire spectral range for all gratings: H-high band ($\lambda 1.538 - $

$1.678 \mu m$), K-long band ($\lambda 2.199 - 2.400 \ \mu m$). We derived the rotational velocity $V$ and velocity dispersion $\sigma_\star$ as the first and second moment from pPXF fitting. We then calculated the $V_{rms} = \sqrt{V^2 + \sigma_\star^2}$. During the fitting process, we also accounted for instrument broadening by convolving the template spectra with a constant dispersion from HSIM.

**Figure 2** shows the fitting results for the central bins of mock cubes without BH. The vertical black-dashed lines represent important stellar absorption features used in the pPXF fitting. The black lines represent the data, and the red lines are the best-fitting models. The residuals (data – model) are depicted as green points.

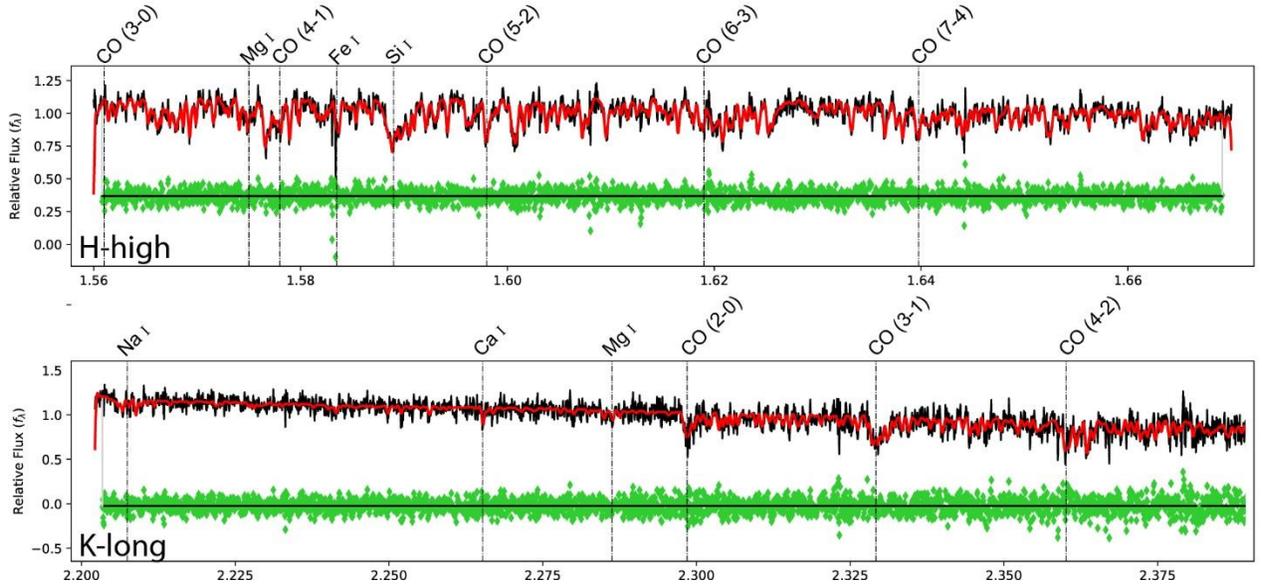

Figure 2. Spectrum fitting for central bin of HSIM gratings

**Figure 3** present the kinematic maps derived from HSIM cubes using pPXF procedure of (left) H-band and (right) K-long band, separated by a black vertical line. In each band, we show row panels corresponding to two BH mass scenarios: (top) without BH and (bottom) with $M_{\rm BH} = 5\% M_{NSC} = 2.5 \times 10^5 M_\odot$. These maps include rotational velocity ($V$), velocity dispersion ($\sigma_\star$), and root-mean-squared velocity ($V_{\rm rms}$) from left to right, respectively.

In all case, the galaxy indicates a slow rotation with $V \lesssim 5$ km.s$^{-1}$ and the $\sigma_\star$ increase from 20 – 30 km.s$^{-1}$ with radius from 0.1 – 0.8 arcsec. This is consistent to the previous kinematics studies using WHT data [48]. Notably, the diferrent appear at the center where BH dominates the gravitational potential, marked by the red circle. In case of no BH, there is a significant drop in $\sigma_\star$. This feature is predicted by [75] in the galaxy with a small or no BH with a wide range of Sersic index and power-law indices of core-Sersic profile. In work of [41], this feature is desmonstrated numerically and also in kinematics simulation. In contrast, a massive BH existence cause the peak up in $\sigma_\star$ profile due to its superior contribution in total gravitaional at this scale. We can see this clearly in case of $M_{BH} = 2.5 \times 10^5 M_\odot$ for VCC 1861 with a raise of $\sigma_\star$ up to 24 km.s$^{-1}$ within 0.1" form center. This indicates a gravitational contribution of IMBH to the kinematic profile that can be clearly distinguished.

Additionally, stellar kinematics extracted from the H-high and K-long bands show a high agreement, with differences below 5%. The H-high band exhibits a higher S/N due to its higher flux intensity.

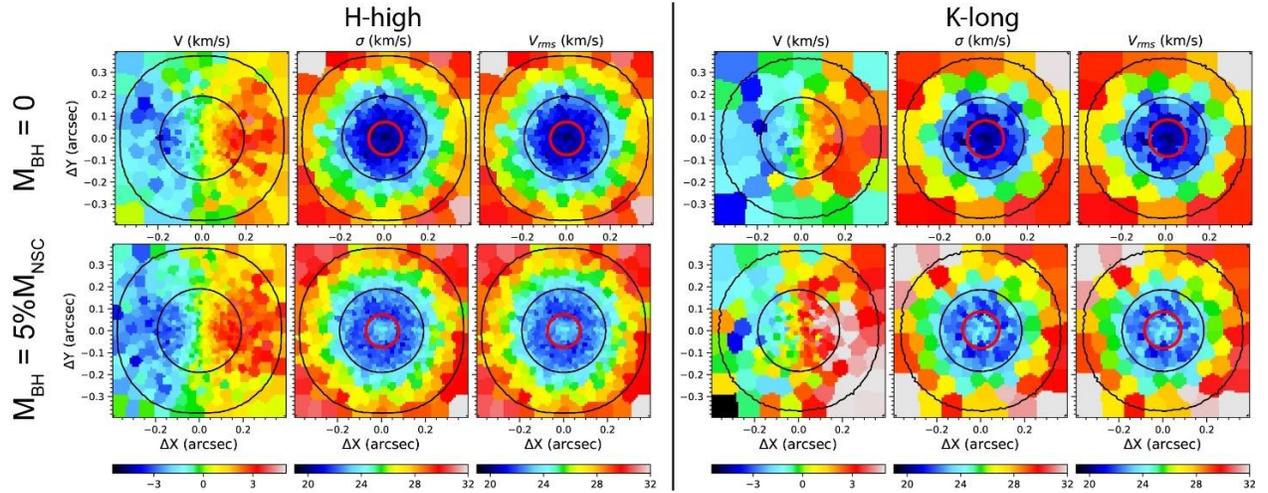

**Figure 3**. Kinematic maps extracted from HSIM cube in high spectral resolution (H-high band & K-long band) with two scenarios: (top) no BH and (bottom) MBH = 5%M$_{NSC}$

## 4.2. Bayesian inference

The V$_{rms}$ profile derived from pPXF fitting is used for BH mass recovery. We apply a Bayesian inference framework to derive the best-fitting JAM$_{cyl}$ parameters utiliztng Adaptive Metropolish (adamet) algorithm [76]. For the data and model comparison, we calculate the chi-squared factor $\chi^2$ which given by $\chi^2 = \frac{1}{\sigma^2}\Sigma_i^n(V_{rms,i} - \bar{V}_{rms,i})$, where $V_{rms,i}$ and $\bar{V}_{rms,i}$ is root-mean-square velocity of data and JAM$_{cyl}$ model, respectively. We use four free parameters in JAM$_{cyl}$ momel, including inclination $i^0$, BH mass in logarithmic scale $log(M_{BH})$, mass-to-light ratio in HST/WFC F850LP filter $M/L_{F850LP}$ and isotropic $\beta_z$. We adopted a log-space prior for BH mass which can span many orders of magnitude. This chosen allows us to explore the sample space over a wide range of masses. We adopted uniform priors for the other parameters. We performed a total of $10^5$ model running iterations, excluding the first 20% step as a burn-in phase. We tracked the post-burn-in phase (80,000 steps) to generate the posterior distribution functions (PDFs). The best-fitting values were obtained from the medians, and uncertainties were derived from the 3σ of the PDF (corresponding to 0.3 - 99.7% confidence levels).

**Figure 5** presents the adamet MCMC posterior distributions as corner plots. The scatter plot colors indicate the confidence levels, with white representing the highest confidence and black indicating values outside the 3-σ level. The 1-D PDFs are also shown as histograms at the top. The best-fit BH mass and $M/L_{F850LP}$ is close to our input values (see **Table 2** and **Figure 4**). Notably, we can observe an anti-correlation between these parameters in their 2D posterior distribution. It highlighs a common physical feature of their contributions to the total system gravitational potential, necessitating that when one parameter becomes larger, the other becomes smaller. In addition, the $β_z$ is also well-constrained with the difference of $Δβ_z < 0.1$.

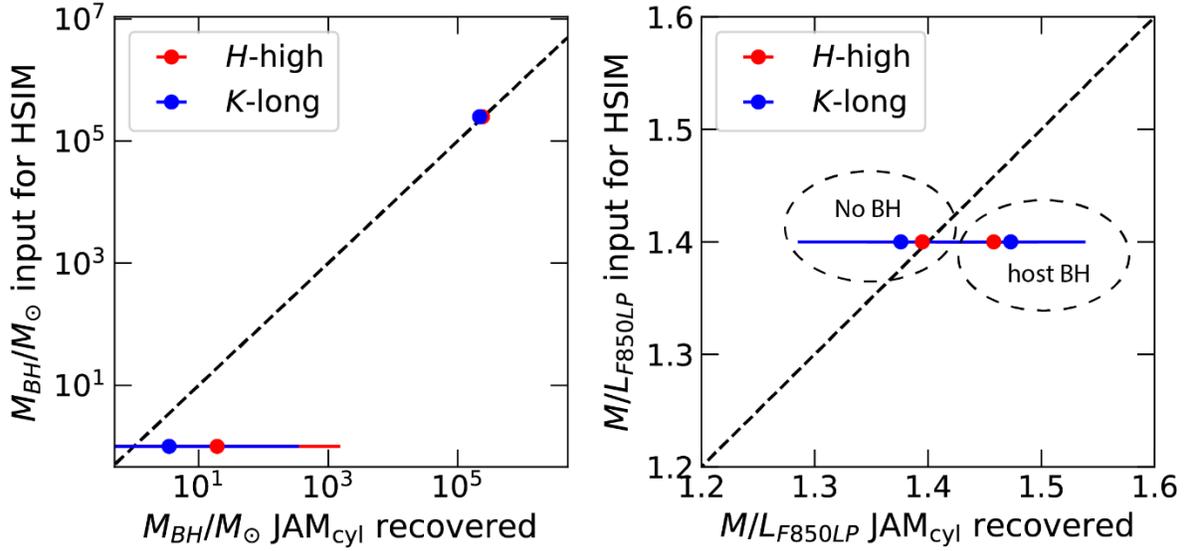

**Figure 4**. Summary of recovery value of two parameters: BH mass and $M/L_{850LP}$

In constrast, the inclinations $i^0$ are not well-constrained, spanning from $30^0 - 90^0$, though it still shows a preference for a best fit at $30^0$, aligning with our input parameters. Our simulation has a FoV of 0.8 arcsec from the center, encompassing only the NSC component. The NSC tend to exhebit an ideal spherical symmetry and appears similar across a wide range of inclinations. It results in poor constraints on the inclination.

Table 2. JAMcyl best-fit values and uncertainties obtained from adamet MCMC process

| Input | Parameters | Priors | Best-fit with $3\sigma$ uncertainties | |
|---|---|---|---|---|
| | | | **H-high** | **K-long** |
| No BH | $i^0$ | 30 – 90 | 37 ± 30 | 34 ± 29 |
| | $log(M_{BH}/M_\odot)$ | 0 – 8 | ≤ 3.184 | ≤ 2.5429 |
| | $M/L_{F850LP}$ | 0 – 8 | 1.395 ± 0.049 | 1.376 ± 0.091 |
| | $\beta_z$ | -1 – 0.99 | 0.010 ± 0.110 | 0.007 ± 0.240 |
| $log(M_{BH}/M_\odot) = 5.4$ | $i^0$ | 30 – 90 | 33 ± 29 | 31 ± 30 |
| | $log(M_{BH}/M_\odot)$ | 0 – 8 | 5.382 ± 0.049 | 5.337 ± 0.090 |
| | $M/L_{F850LP}$ | 0 – 8 | 1.458 ± 0.039 | 1.473 ± 0.066 |
| | $\beta_z$ | -1 – 0.99 | 0.014 ± 0.063 | 0.014 ± 0.100 |

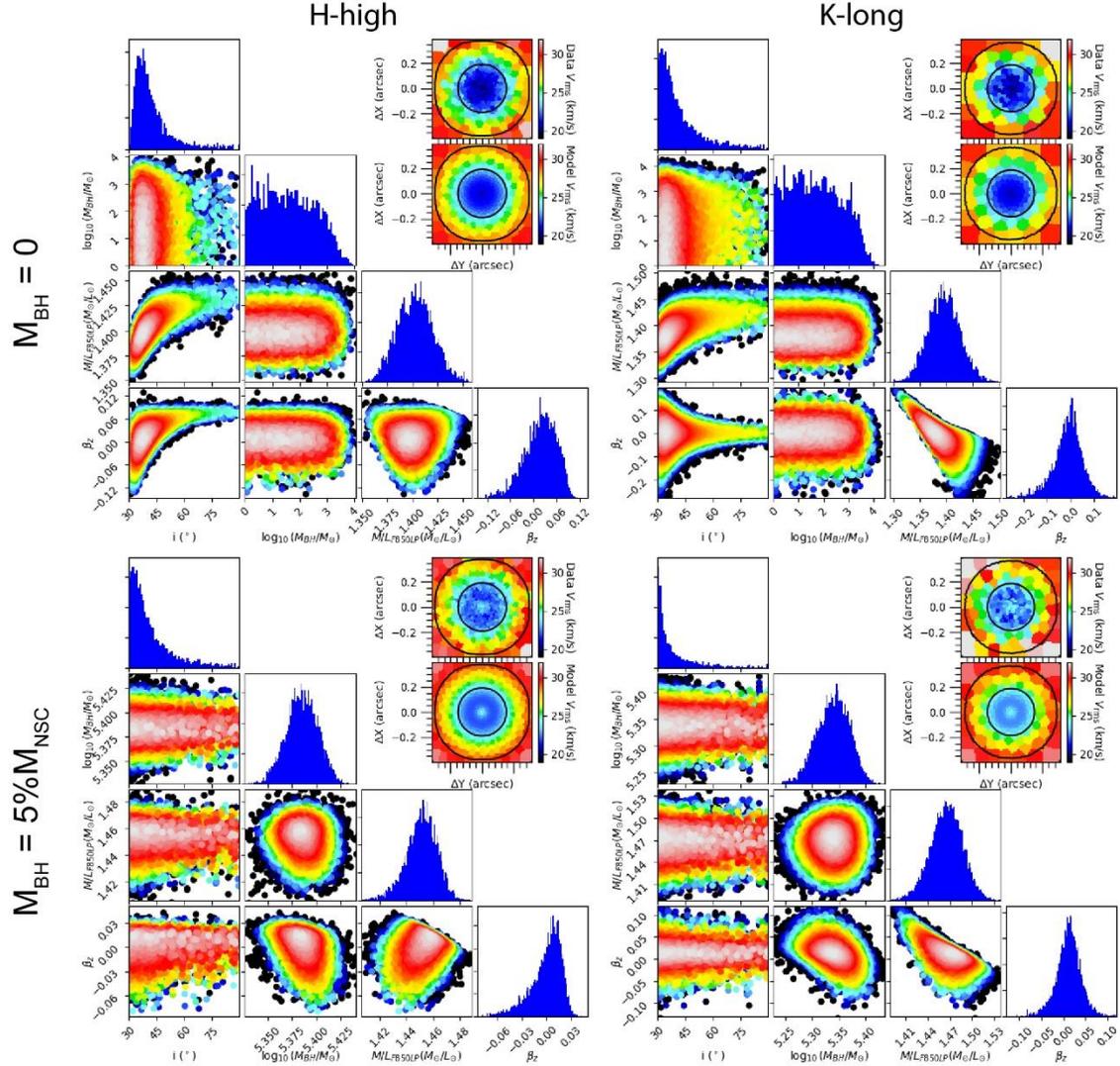

**Figure 5**. Corner plots show the MCMC reults for medium resolution gratings: **(left)** H-band and **(right)** K-band with two different scenarios: **(upper)** no BH and **(lower)** a central IMBH with mass of 5% NSC mass. The top histograms show each parameter 1D posterior distribution, with 2D PDF shown below. The colors from white to black depict different confident levels. The data are compared to best-fitting model at top-right corner.

## 5. Conclusion

These IMBHs are crucial to knowledge of the formation mechanism and evolution of SMBHs. In this upcoming future, ELT promises to improve the detection of these objects. In this study, we have focused on the ability of ELT in IMBH detecting with galaxies in Virgo Cluster (distance of $D \approx 16.5$ Mpc). This aims to expand the survey further than the previous IMBH survey [41].

The high-spectral resolution gratings of ELT, including H-high and K-long at 10 mas spatial resolution scale, have abiliy to detect IMBHs with mass of $M_{BH} = 2.5 \times 10^5 M_\odot$ at the distance of 16.5 Mpc.

**Acknowledgements**

This research is funded by University of Science, VNU-HCM under grant number T2023-105.

Facilities: HST

Software: emcee, python 3.12: https://www.python.org/, Matplotlib 3.6.0: https://matplotlib.org/, numpy 1.22: https://www.scipy.org/install.html, scipy 1.3.1: https://www.scipy.org/install.html, photutils 0.7: https://photutils.readthedocs.io/en/stable//, astropy 5.1: (Astropy Collaboration et al. 2022), plotbin 3.1.3: https://pypi.org/project/plotbin/, MPFIT: http://purl.com/net/mpfit, adamet 2.0.9: (Cappellari et al. 2013), jampy 7.2.5 (Cappellari 2020), pPXF 9.2.2 (Cappellari 2023), vorbin 3.1.5 (Cappellari & Copin 2003), MgeFit 5.0.14 (Cappellari 2002), and HSIM 3.11 (Zieleniewski et al. 2015).

**Literature - References**

[1] J. ~R. Oppenheimer and H. Snyder, "On Continued Gravitational Contraction," *Phys. Rev.*, vol. 56, no. 5, pp. 455–459, Sep. 1939, doi: 10.1103/PhysRev.56.455.

[2] A. J. Barth, P. Martini, C. H. Nelson, and L. C. Ho, "Iron Emission in the z = 6.4 Quasar SDSS J114816.64+525150.3," *Astrophys. J.*, vol. 594, no. 2, p. L95, Aug. 2003, doi: 10.1086/378735.

[3] D. J. Mortlock et al., "A luminous quasar at a redshift of z = 7.085," *Nature*, vol. 474, no. 7353, pp. 616–619, Jun. 2011, doi: 10.1038/nature10159.

[4] T. Izumi et al., "Subaru High-z Exploration of Low-luminosity Quasars (SHELLQs). XIII. Large-scale Feedback and Star Formation in a Low-luminosity Quasar at z = 7.07 on the Local Black Hole to Host Mass Relation," *Astrophys. J.*, vol. 914, no. 1, p. 36, Jun. 2021, doi: 10.3847/1538-4357/abf6dc.

[5] A.-C. Eilers et al., "EIGER. III. JWST/NIRCam Observations of the Ultraluminous High-redshift Quasar J0100+2802," *Astrophys. J.*, vol. 950, no. 1, p. 68, Jun. 2023, doi: 10.3847/1538-4357/acd776.

[6] R. Endsley et al., "ALMA confirmation of an obscured hyperluminous radio-loud AGN at z = 6.853 associated with a dusty starburst in the 1.5 deg2 COSMOS field," *Mon. Not. R. Astron. Soc.*, vol. 520, no. 3, pp. 4609–4620, 2023, doi: 10.1093/mnras/stad266.

[7] R. Maiolino et al., "A small and vigorous black hole in the early Universe," *Nature*, pp. 1–3, 2024.

[8] S. van Wassenhove, M. Volonteri, M. ~G. Walker, and J. ~R. Gair, "Massive black holes lurking in Milky Way satellites," *Monthly Notices of the Royal Astronomical*

*Society*, vol. 408, no. 2, pp. 1139–1146, Oct. 2010, doi: 10.1111/j.1365-2966.2010.17189.x.

[9] M. Volonteri, "The Formation and Evolution of Massive Black Holes," *Science (80-. ).*, vol. 337, no. 6094, p. 544, Aug. 2012, doi: 10.1126/science.1220843.

[10] M. C. Miller and M. B. Davies, "An Upper Limit to the Velocity Dispersion of Relaxed Stellar Systems without Massive Black Holes," *Astrophys. J.*, vol. 755, no. 1, p. 81, Aug. 2012, doi: 10.1088/0004-637X/755/1/81.

[11] M. Giersz, N. Leigh, A. Hypki, N. Lützgendorf, and A. Askar, "MOCCA code for star cluster simulations - IV. A new scenario for intermediate mass black hole formation in globular clusters," *Monthly Notices of the Royal Astronomical Society*, vol. 454, no. 3, pp. 3150–3165, Dec. 2015, doi: 10.1093/mnras/stv2162.

[12] M. C. Begelman and M. J. Rees, "The fate of dense stellar systems," *Mon. Not. R. Astron. Soc.*, vol. 185, no. 4, pp. 847–860, 1978, doi: 10.1093/mnras/185.4.847.

[13] E. ~K. Verolme *et al.*, "A SAURON study of M32: measuring the intrinsic flattening and the central black hole mass," *Monthly Notices of the Royal Astronomical Society*, vol. 335, no. 3, pp. 517–525, Sep. 2002, doi: 10.1046/j.1365-8711.2002.05664.x.

[14] J. E. Greene, "Low-mass black holes as the remnants of primordial black hole formation," *Nat. Commun.*, vol. 3, p. 1304, Dec. 2012, doi: 10.1038/ncomms2314.

[15] S. Bonoli, L. Mayer, and S. Callegari, "Massive black hole seeds born via direct gas collapse in galaxy mergers: their properties, statistics and environment," *Monthly Notices of the Royal Astronomical Society*, vol. 437, no. 2, pp. 1576–1592, Jan. 2014, doi: 10.1093/mnras/stt1990.

[16] A. E. Reines, G. R. Sivakoff, K. E. Johnson, and C. L. Brogan, "An actively accreting massive black hole in the dwarf starburst galaxy Henize 2-10," *Nature*, vol. 470, no. 7332, pp. 66–68, 2011, doi: 10.1038/nature09724.

[17] M. Mezcua, F. Civano, G. Fabbiano, T. Miyaji, and S. Marchesi, "A POPULATION OF INTERMEDIATE-MASS BLACK HOLES IN DWARF STARBURST GALAXIES UP TO REDSHIFT = 1.5," *Astrophys. J.*, vol. 817, no. 1, p. 20, Jan. 2016, doi: 10.3847/0004-637X/817/1/20.

[18] V. ~F. Baldassare, A. ~E. Reines, E. Gallo, and J. ~E. Greene, "A ~50,000 M$_{\odot}$ Solar Mass Black Hole in the Nucleus of RGG 118," *Astrophys. J.l*, vol. 809, p. L14, Aug. 2015, doi: 10.1088/2041-8205/809/1/L14.

[19] S. F. Portegies Zwart and S. L. W. McMillan, "The Runaway Growth of Intermediate-Mass Black Holes in Dense Star Clusters," *Astrophys. J.*, vol. 576, no. 2, pp. 899–907, 2002, doi: 10.1086/341798.

[20] S. F. Portegies Zwart, H. Baumgardt, P. Hut, J. Makino, and S. L. W. McMillan, "Formation of massive black holes through runaway collisions in dense young star clusters," *Nature*, vol. 428, no. 6984, pp. 724–726, 2004, doi: 10.1038/nature02448.

[21] K. Gebhardt, R. ~M. Rich, and L. C. Ho, "An Intermediate-Mass Black Hole in the Globular Cluster G1: Improved Significance from New Keck and Hubble Space Telescope Observations," *Astrophys. J.*, vol. 634, no. 2, pp. 1093–1102, Dec. 2005, doi: 10.1086/497023.

[22] B. Kızıltan, H. Baumgardt, and A. Loeb, "An intermediate-mass black hole in the centre of the globular cluster 47 Tucanae," *Nature*, vol. 542, no. 7640, pp. 203–205, 2017, doi: 10.1038/nature21361.

[23] K. Gebhardt *et al.*, "M33: A Galaxy with No Supermassive Black Hole," *Astron. J.*, vol.

[24] M. Valluri, L. Ferrarese, D. Merritt, and C. L. Joseph, "The Low End of the Supermassive Black Hole Mass Function: Constraining the Mass of a Nuclear Black Hole in NGC 205 via Stellar Kinematics," *Astrophys. J.*, vol. 628, no. 1, pp. 137–152, 2005, doi: 10.1086/430752.

[25] L. J. Latimer, A. E. Reines, A. Bogdan, and R. Kraft, "The AGN Fraction in Dwarf Galaxies from eROSITA: First Results and Future Prospects," *Astrophys. J. Lett.*, vol. 922, no. 2, p. L40, Dec. 2021, doi: 10.3847/2041-8213/ac3af6.

[26] K. L. Birchall, M. G. Watson, and J. Aird, "X-ray detected AGN in SDSS dwarf galaxies," *Mon. Not. R. Astron. Soc.*, vol. 492, no. 2, pp. 2268–2284, 2020, doi: 10.1093/mnras/staa040.

[27] S. M. Lemons, A. E. Reines, R. M. Plotkin, E. Gallo, and J. E. Greene, "AN X-RAY-SELECTED SAMPLE OF CANDIDATE BLACK HOLES IN DWARF GALAXIES," *Astrophys. J.*, vol. 805, no. 1, p. 12, May 2015, doi: 10.1088/0004-637X/805/1/12.

[28] A. Rau et al., "Exploring the Optical Transient Sky with the Palomar Transient Factory," *Publ. Astron. Soc. Pacific*, vol. 121, no. 886, p. 1334, Oct. 2009, doi: 10.1086/605911.

[29] N. M. Law et al., "The Palomar Transient Factory: System Overview, Performance, and First Results," *Publ. Astron. Soc. Pacific*, vol. 121, no. 886, p. 1395, Nov. 2009, doi: 10.1086/648598.

[30] M. J. Graham et al., "The Zwicky Transient Facility: Science Objectives," *Publ. Astron. Soc. Pacific*, vol. 131, no. 1001, p. 78001, May 2019, doi: 10.1088/1538-3873/ab006c.

[31] K. P. Mooley et al., "THE CALTECH-NRAO STRIPE 82 SURVEY (CNSS) PAPER. I. THE PILOT RADIO TRANSIENT SURVEY IN 50 DEG2," *Astrophys. J.*, vol. 818, no. 2, p. 105, Feb. 2016, doi: 10.3847/0004-637X/818/2/105.

[32] A. J. Barth, L. E. Strigari, M. C. Bentz, J. E. Greene, and L. C. Ho, "Dynamical Constraints on the Masses of the Nuclear Star Cluster and Black Hole in the Late-Type Spiral Galaxy NGC 3621," *Astrophys. J.*, vol. 690, no. 1, pp. 1031–1044, Jan. 2009, doi: 10.1088/0004-637X/690/1/1031.

[33] M. Den Brok et al., "Measuring the mass of the central black hole in the bulgeless galaxy NGC 4395 from gas dynamical modeling," *Astrophys. J.*, vol. 809, no. 1, p. 101, 2015, doi: 10.1088/0004-637X/809/1/101.

[34] D. D. Nguyen et al., "Nearby Early-type Galactic Nuclei at High Resolution: Dynamical Black Hole and Nuclear Star Cluster Mass Measurements," *Astrophys. J.*, vol. 858, no. 2, p. 118, May 2018, doi: 10.3847/1538-4357/aabe28.

[35] D. D. Nguyen et al., "Improved Dynamical Constraints on the Masses of the Central Black Holes in Nearby Low-mass Early-type Galactic Nuclei and the First Black Hole Determination for NGC 205," *Astrophys. J.*, vol. 872, no. 1, p. 104, Feb. 2019, doi: 10.3847/1538-4357/aafe7a.

[36] A. C. Seth et al., "The NGC 404 Nucleus: Star Cluster and Possible Intermediate-mass Black Hole," *Astrophys. J.*, vol. 714, no. 1, pp. 713–731, May 2010, doi: 10.1088/0004-637X/714/1/713.

[37] D. D. Nguyen et al., "Improved Dynamical Constraints on the Mass of the Central Black Hole in NGC 404," *Astrophys. J.*, vol. 836, no. 2, p. 237, Feb. 2017, doi: 10.3847/1538-4357/aa5cb4.

[38] T. A. Davis et al., "Revealing the intermediate-mass black hole at the heart of the dwarf

[38] galaxy NGC 404 with sub-parsec resolution ALMA observations," *Monthly Notices of the Royal Astronomical Society*, vol. 496, no. 4, pp. 4061–4078, Jul. 2020, doi: 10.1093/mnras/staa1567.

[39] M. Häberle *et al.*, "Fast-moving stars around an intermediate-mass black hole in {\ensuremath{\omega}} Centauri," *Nature*, vol. 631, no. 8020, pp. 285–288, Jul. 2024, doi: 10.1038/s41586-024-07511-z.

[40] R. Pechetti *et al.*, "Detection of a 100,000 M $_{{\ensuremath{\odot}}}$ black hole in M31's Most Massive Globular Cluster: A Tidally Stripped Nucleus," *Astrophys. J.*, vol. 924, no. 2, p. 48, Jan. 2022, doi: 10.3847/1538-4357/ac339f.

[41] D. D. Nguyen *et al.*, "Simulating intermediate black hole mass measurements for a sample of galaxies with nuclear star clusters using ELT/HARMONI high spatial resolution integral-field stellar kinematics," *arXiv e-prints*, p. arXiv:2408.00239, Jul. 2024, doi: 10.48550/arXiv.2408.00239.

[42] B. Binggeli, G. ~A. Tammann, and A. Sandage, "Studies of the Virgo Cluster. VI. Morphological and Kinematical Structure of the Virgo Cluster," \aj, vol. 94, p. 251, Aug. 1987, doi: 10.1086/114467.

[43] M. Volonteri, F. Haardt, and K. Gültekin, "Compact massive objects in Virgo galaxies: the black hole population," *Mon. Not. R. Astron. Soc.*, vol. 384, no. 4, pp. 1387–1392, 2008, doi: 10.1111/j.1365-2966.2008.12911.x.

[44] A. W. Graham, R. Soria, and B. L. Davis, "Expected intermediate-mass black holes in the Virgo cluster – II. Late-type galaxies," *Mon. Not. R. Astron. Soc.*, vol. 484, no. 1, pp. 814–831, 2018, doi: 10.1093/mnras/sty3068.

[45] H. C. Ferguson, "Galaxy Populations in the Fornax and Virgo Clusters," \apss, vol. 157, no. 1–2, pp. 227–233, Jul. 1989, doi: 10.1007/BF00637334.

[46] C. Spengler *et al.*, "Virgo Redux: The Masses and Stellar Content of Nuclei in Early-type Galaxies from Multiband Photometry and Spectroscopy," *Astrophys. J.*, vol. 849, no. 1, p. 55, Nov. 2017, doi: 10.3847/1538-4357/aa8a78.

[47] N. Scott and A. W. Graham, "UPDATED MASS SCALING RELATIONS FOR NUCLEAR STAR CLUSTERS AND A COMPARISON TO SUPERMASSIVE BLACK HOLES," *Astrophys. J.*, vol. 763, no. 2, p. 76, Jan. 2013, doi: 10.1088/0004-637X/763/2/76.

[48] E. Toloba *et al.*, "STELLAR KINEMATICS AND STRUCTURAL PROPERTIES OF VIRGO CLUSTER DWARF EARLY-TYPE GALAXIES FROM THE SMAKCED PROJECT. II. THE SURVEY AND A SYSTEMATIC ANALYSIS OF KINEMATIC ANOMALIES AND ASYMMETRIES," *Astrophys. J. Suppl. Ser.*, vol. 215, no. 2, p. 17, Dec. 2014, doi: 10.1088/0067-0049/215/2/17.

[49] A. Sybilska *et al.*, "The hELENa project – I. Stellar populations of early-type galaxies linked with local environment and galaxy mass," *Mon. Not. R. Astron. Soc.*, vol. 470, no. 1, pp. 815–838, 2017, doi: 10.1093/mnras/stx1138.

[50] B. S. Ryden, D. M. Terndrup, R. W. Pogge, and T. R. Lauer, "Detailed Surface Photometry of Dwarf Elliptical and Dwarf S0 Galaxies in the Virgo Cluster," *Astrophys. J.*, vol. 517, no. 2, p. 650, Jun. 1999, doi: 10.1086/307201.

[51] P. Côté *et al.*, "The ACS Virgo Cluster Survey. VIII. The Nuclei of Early-Type Galaxies," *Astrophys. J.s*, vol. 165, no. 1, pp. 57–94, Jul. 2006, doi: 10.1086/504042.

[52] S. Zieleniewski *et al.*, "hsim: a simulation pipeline for the HARMONI integral field spectrograph on the European ELT," *Mon. Not. R. Astron. Soc.*, vol. 453, no. 4, pp.

3754–3765, Nov. 2015, doi: 10.1093/mnras/stv1860.

[53] J. Krist, "Simulation of HST PSFs using Tiny Tim," in *Astronomical Data Analysis Software and Systems IV*, Jan. 1995, vol. 77, p. 349.

[54] J. E. Krist, R. N. Hook, and F. Stoehr, "20 years of Hubble Space Telescope optical modeling using Tiny Tim," in *Optical Modeling and Performance Predictions V*, Oct. 2011, vol. 8127, p. 81270J, doi: 10.1117/12.892762.

[55] R. J. Avila, W. ~J. Hack, and STScI AstroDrizzle Team, "AstroDrizzle: Aligning Images From Multiple Instruments," in *American Astronomical Society Meeting Abstracts \#220*, May 2012, vol. 220, p. 135.13.

[56] R. I. Jedrzejewski, "CCD surface photometry of elliptical galaxies - I. Observations, reduction and results.," *Monthly Notices of the Royal Astronomical Society*, vol. 226, pp. 747–768, Jun. 1987, doi: 10.1093/mnras/226.4.747.

[57] A. W. Graham, P. Erwin, I. Trujillo, and A. Asensio Ramos, "A New Empirical Model for the Structural Analysis of Early-Type Galaxies, and A Critical Review of the Nuker Model," \aj, vol. 125, pp. 2951–2963, Jun. 2003, doi: 10.1086/375320.

[58] I. Trujillo, P. Erwin, A. Asensio Ramos, and A. W. Graham, "Evidence for a New Elliptical-Galaxy Paradigm: Sérsic and Core Galaxies," \aj, vol. 127, pp. 1917–1942, Apr. 2004, doi: 10.1086/382712.

[59] J. L. Sersic, *Atlas de galaxias australes*. Córdoba: Obs. Astron. Univ. Nacional de Córdoba, 1968.

[60] C. ~B. Markwardt, "Non-linear Least-squares Fitting in IDL with MPFIT," in *Astronomical Data Analysis Software and Systems XVIII*, Sep. 2009, vol. 411, p. 251, doi: 10.48550/arXiv.0902.2850.

[61] J. van de Sande, M. Kriek, M. Franx, R. Bezanson, and P. G. van Dokkum, "The Relation between Dynamical Mass-to-light Ratio and Color for Massive Quiescent Galaxies out to z \raisebox{-0.5ex}\textasciitilde 2 and Comparison with Stellar Population Synthesis Models," *Astrophys. J.*, vol. 799, no. 2, p. 125, Feb. 2015, doi: 10.1088/0004-637X/799/2/125.

[62] C. Maraston and G. Strömbäck, "Stellar population models at high spectral resolution," *Monthly Notices of the Royal Astronomical Society*, vol. 418, no. 4, pp. 2785–2811, Dec. 2011, doi: 10.1111/j.1365-2966.2011.19738.x.

[63] B. Gustafsson, B. Edvardsson, K. Eriksson, U. ~G. Jørgensen, \rA. Nordlund, and B. Plez, "A grid of MARCS model atmospheres for late-type stars. I. Methods and general properties," \aap, vol. 486, no. 3, pp. 951–970, Aug. 2008, doi: 10.1051/0004-6361:200809724.

[64] A. Seth, M. Agüeros, D. Lee, and A. Basu-Zych, "The Coincidence of Nuclear Star Clusters and Active Galactic Nuclei," *Astrophys. J.*, vol. 678, no. 1, pp. 116–130, May 2008, doi: 10.1086/528955.

[65] M. Cappellari, "Measuring the inclination and mass-to-light ratio of axisymmetric galaxies via anisotropic Jeans models of stellar kinematics," *Mon. Not. R. Astron. Soc.*, vol. 390, no. 1, pp. 71–86, Oct. 2008, doi: 10.1111/j.1365-2966.2008.13754.x.

[66] M. Cappellari, "Efficient solution of the anisotropic spherically aligned axisymmetric Jeans equations of stellar hydrodynamics for galactic dynamics," *Mon. Not. R. Astron. Soc.*, vol. 494, no. 4, pp. 4819–4837, Jun. 2020, doi: 10.1093/mnras/staa959.

[67] M. Lipka *et al.*, "The VIRUS-dE Survey I: Stars in dwarf elliptical galaxies - 3D dynamics and radially resolved stellar initial mass functions." 2024, [Online]. Available:

https://arxiv.org/abs/2409.10518.
[68] N. A. Thatte *et al.*, "The E-ELT first light spectrograph HARMONI: capabilities and modes," in *Ground-based and Airborne Instrumentation for Astronomy VI*, Aug. 2016, vol. 9908, p. 99081X, doi: 10.1117/12.2230629.
[69] D. D. Nguyen, M. Cappellari, and M. Pereira-Santaella, "Simulating supermassive black hole mass measurements for a sample of ultramassive galaxies using ELT/HARMONI high-spatial-resolution integral-field stellar kinematics," *Monthly Notices of the Royal Astronomical Society*, vol. 526, no. 3, pp. 3548–3569, Dec. 2023, doi: 10.1093/mnras/stad2860.
[70] S. Kendrew *et al.*, "Simulated stellar kinematics studies of high-redshift galaxies with the HARMONI Integral Field Spectrograph," *Mon. Not. R. Astron. Soc.*, vol. 458, no. 3, pp. 2405–2422, 2016, doi: 10.1093/mnras/stw438.
[71] J. ~K. Kotilainen, T. Hyvönen, J. Reunanen, and V. ~D. Ivanov, "Near-infrared spectroscopy of stellar populations in nearby spiral galaxies," *Monthly Notices of the Royal Astronomical Society*, vol. 425, no. 2, pp. 1057–1067, Sep. 2012, doi: 10.1111/j.1365-2966.2012.21425.x.
[72] M. Lyubenova, H. Kuntschner, M. Rejkuba, D. ~R. Silva, M. Kissler-Patig, and L. ~E. Tacconi-Garman, "Integrated J- and H-band spectra of globular clusters in the LMC: implications for stellar population models and galaxy age dating," *\aap*, vol. 543, p. A75, Jul. 2012, doi: 10.1051/0004-6361/201218847.
[73] M. Cappellari and Y. Copin, "Adaptive spatial binning of integral-field spectroscopic data using Voronoi tessellations," *Monthly Notices of the Royal Astronomical Society*, vol. 342, no. 2, pp. 345–354, Jun. 2003, doi: 10.1046/j.1365-8711.2003.06541.x.
[74] M. Cappellari, "Full spectrum fitting with photometry in PPXF: stellar population versus dynamical masses, non-parametric star formation history and metallicity for 3200 LEGA-C galaxies at redshift z {\ensuremath{\approx}} 0.8," *Monthly Notices of the Royal Astronomical Society*, vol. 526, no. 3, pp. 3273–3300, Dec. 2023, doi: 10.1093/mnras/stad2597.
[75] S. Tremaine *et al.*, "A Family of Models for Spherical Stellar Systems," *Astron. J.*, vol. 107, p. 634, Feb. 1994, doi: 10.1086/116883.
[76] M. Cappellari *et al.*, "The ATLAS3D project – XV. Benchmark for early-type galaxies scaling relations from 260 dynamical models: mass-to-light ratio, dark matter, Fundamental Plane and Mass Plane," *Mon. Not. R. Astron. Soc.*, vol. 432, no. 3, pp. 1709–1741, Jul. 2013, doi: 10.1093/mnras/stt562.